\documentclass[11pt,onecolumn]{article}
\usepackage{times}
\usepackage{amsmath}
\usepackage{amssymb}
\usepackage{paralist}
\usepackage{boxedminipage}
\usepackage{epsfig,graphics}
\usepackage[table]{xcolor}
\usepackage{comment}
\usepackage{color}
\usepackage{enumitem}
\setlist{topsep=1pt,itemsep=0pt,parsep=1pt,itemindent=0pt,leftmargin=2em}
\usepackage{wrapfig}

\usepackage{listings}
\definecolor{dgray}{gray}{0.2}
\lstdefinestyle{cosc}{language=C,
keywordstyle=\bfseries\color{blue},
commentstyle=\color{magenta},
stringstyle=\color{green},
basicstyle=\color{black}\scriptsize,
frame=single,
mathescape
}

\usepackage[hyphens]{url}
\usepackage[hyphenbreaks]{breakurl}
\usepackage{hyperref}
\hypersetup{
  colorlinks=true,
  citecolor=blue,
  linkcolor=purple,
  urlcolor=black,
  citebordercolor={1 1 1},
  linkbordercolor={1 1 1},
}
\usepackage{bibentry}

\usepackage[hmargin=1in,vmargin=1in]{geometry}

\newcommand{\minihead}[1]{\vspace{0.1em}\noindent{\bf #1}}

\title{Towards Designing A Secure Plausibly Deniable System for Mobile Devices against Multi-snapshot Adversaries - A Preliminary Design}
\author{Bo Chen\\Department of Computer Science\\Michigan Technological University\\bchen@mtu.edu}
\date{}

\begin{document}

\newcommand{\nwf}[1]{{\color{red}  {\bf NWF Comment}: #1}}

\thispagestyle{empty}

\newpage
\setcounter{page}{1}

\maketitle
\begin{abstract}
Mobile computing devices have been used broadly to store, manage and process sensitive or even mission critical data. To protect confidentiality of data stored in mobile devices, major mobile operating systems use full disk encryption, which relies on traditional encryption mechanisms and requires that decryption keys will not be disclosed. This however, is not necessarily true, since an active attacker may coerce victims for decryption keys. Plausibly deniable encryption (PDE) can defend against such a coercive attacker by disguising the true secret key with a decoy key. Leveraging concept of PDE, various deniable storage systems have been built for both PC and mobile platforms. However, a secure PDE system for mobile devices is still missing which can be compatible with mainstream mobile devices and, meanwhile, remains secure when facing a strong multi-snapshot adversary.   

In this work, we propose a preliminary PDE system design for mobile computing devices using flash memory as underlying storage medium. Ours is the first secure PDE system for mobile devices which has the following new design features: 1) it is compatible with mainstream mobile devices due to its integration of PDE into flash translation layer (FTL), the most popular form of flash memory being used by modern mobile devices; and 2) it can defend against the multi-snapshot adversary by denying hidden writes (over the flash memory) caused by hidden sensitive data using random dummy writes. 
\end{abstract}
\section{Introduction}
Nowadays, more and more people use mobile computing devices (e.g., smartphones, tablets, smart home assistants like Amazon Echo and Google Home) for social networking, e-commerce, online banking, etc. This brings a lot of convenience, but also creates a significant security concern on the mobile computing devices which have been used to store, manage, and process sensitive personal/work data. Especially, users from federal agencies or even military departments increasingly turn to mobile devices for handling mission critical data in their daily operations. This makes it an urgent need to protect sensitive data stored in mobile devices. To protect confidentiality of sensitive data, major mobile operating systems have incorporated a certain level of encryption. For example, Android has supported full disk encryption since version 3.0~\cite{androidfde}. Traditional encryption mechanisms, including both symmetric encryption and asymmetric encryption, however, may remain secure if the key for decrypting ciphertexts remains secret. However, an emerging attack is a coercive attack, in which the victim of a mobile device is forced to disclose the decryption key. For example, a human rights worker collects criminal evidence using her smart phone in a region of oppression; to keep the data secret, she stores them encrypted via an encryption tool;, when she tries to cross border, the inspector notices existence of encrypted data and may coerce her to disclose the decryption key. Plausibly Deniable Encryption (PDE) has been proposed to defend against such adversaries who can coerce users into revealing the secret keys. PDE works as follows: Original sensitive data are encrypted into ciphertexts in such a way that, when using a decoy key, reasonable and innocuous plaintexts will be generated; only when using a true key, the original sensitive data will be obtained; therefore upon coerced, the victim can disclose the decoy key and, the adversary uses the decoy key to decrypt the ciphertexts, obtaining the innocuous plaintexts rather than the original sensitive data.

Instantiating PDE in cryptography could be a challenging problem. Instead, a variety of deniable storage systems have been built to provide deniability, by leveraging concept of PDE. This includes TrueCrypt~\cite{truecrypt}, Rubberhose~\cite{Rubberhose}, HIVE~\cite{blass2014toward}, Gracewipe~\cite{blass2014toward}, Steganographic File Systems~\cite{stegfs_IH1998,stegfs_IH1999,stegfs_ICDE2003}, etc. These existing PDE systems were originally built for PC, but they are not directly applicable to the ever-growing mobile computing devices, because: First, compared to a PC platform, a mobile platform is usually equipped with limited computational resources and sensitive to energy consumption. The existing PDE systems~\cite{truecrypt,Rubberhose,stegfs_IH1998,stegfs_IH1999,stegfs_ICDE2003,blass2014toward,blass2014toward} may cause large overhead and are thus not suitable for mobile devices. Second, mobile computing devices usually use flash memory storage rather than mechanical drives. Compared to the mechanical drives, flash memory has significantly different hardware features, e.g., unable to support in-place updates, vulnerable to wear. The existing PDE systems~\cite{truecrypt,Rubberhose,stegfs_IH1998,stegfs_IH1999,stegfs_ICDE2003,blass2014toward,blass2014toward} were originally built for mechanical disks, and might neglect the special nature of flash memory, causing various issues in both performance and security.

Recently, PDE systems for mobile platforms have been proposed~\cite{chang2018user,mobipluto_acsac2015,chang2018mobiceal,mobiflage_ndss2013,mobiflage_TDSC2014,defy_ndss2015,mobihydra_isc2014}; however, most of them are built on the block layer and unfortunately suffer from deniability compromise in the underlying raw flash memory. A fundamental reason is that raw flash memory requires special internal management (usually implemented in Flash Translation Layer, or FTL) to handle the unique hardware nature of flash memory~\cite{kim2002space}. Even if sensitive data have been isolated carefully on the block layer (and hence remain hidden), they will not be carefully isolated by the ``internal management'' of flash memory which works independently and is transparent to the upper layer. Therefore, ``traces'' of sensitive data may be present in the raw flash memory and observed by the adversaries, compromising deniability~\cite{jia2017ccs}.  
 
Note that the file system or the upper OS cannot ``touch'' the FTL which is part of the flash device hardware, and hence is not able to remove those ``traces''. Therefore, for mobile devices, to prevent deniability compromise, it seems an unavoidable need to move (or partially move) PDE from the block layer to the lower flash memory layer to avoid deniability compromise. DEFTL~\cite{jia2017ccs} takes an initial step to confirm possibility of incorporating PDE into FTL\footnote{Another work DEFY~\cite{defy_ndss2015} also tries to build PDE into flash memory. Their design strongly relies on system properties provided by a flash-specific file system YAFFS, but flash file systems are rarely used in mobile devices now.}. However, DEFTL relies on the hidden volume mechanism (Sec.~\ref{sec:background:hidden}) and is vulnerable to a multi-snapshot adversary, which can have access to the disk multiple times over time and, by comparing different disk snapshots captured, it can detect unaccountable changes on the random data caused by writes performed by the hidden volume, compromising deniability. A PDE system designed for mobile devices which can resist against multi-snapshot adversaries is still missing in the literature. This work thus aims to fill this gap by designing \textit{the first PDE system for mobile devices which can combat multi-snapshot adversaries}. 

\section{Background Knowledge}
\label{sec:background}
\subsection{Flash Memory}
\label{sec:background:flash}

Flash memory is a non-volatile storage medium which can be electrically erased and reprogrammed. Compared to traditional mechanical drives, flash memory can achieve much higher I/O throughput with much lower power consumption, and thus gains popularity in mainstream mobile computing devices like smartphones, tablets, smartwatches, digital camcorders, etc. The flash family contains NAND-type and NOR-type flash, in which NAND type is used broadly as storage medium. NAND flash stores information in an array of memory cells, which are grouped into blocks (a few hundred Kilobytes).
A block is divided into a few pages. Typical page size can be 512 bytes, 2KB, and 4KB. Each page usually contains a small spare out-of-band (OOB) area which is used to store various metadata (e.g., error correction code). 
Compared to mechanical disks, NAND flash has a few unique characteristics. First, it has an erase-before-write design, i.e., a flash cell needs to be erased before it can be overwritten. Second, the unit for a read/program operation is a page, but the unit for an erase operation is a block. Flash memory is update unfriendly since updating a page requires erasing the entire encompassing block. Therefore, it usually uses out-of-place update. Third, a flash block usually has a finite number (e.g., $10$K) of program-erase cycles, and will be worn out if the number of programs/erasures performed over it exceeds a certain threshold.

\minihead{Flash Translation Layer (FTL).} The most popular form of using flash memory is \textit{to emulate it as a block device}. In this way, a flash storage device can be compatible with traditional block file systems like EXT4 and FAT32. Most existing flash storage media (e.g., eMMC cards, MiniSD cards) used in mobile devices are used as block devices. A piece of special firmware, Flash Translation Layer (FTL)~\cite{gupta2009dftl,kang2006superblock,kim2002space,lee2007log,park2008reconfigurable}, is introduced between the file system and the raw NAND flash to transparently handle unique nature of flash memory. FTL usually implements address translation, garbage collection, wear leveling, and bad block management. 
Since flash storage is emulated as a block device, FTL needs to perform \textit{address translation} between block addresses and flash memory addresses. This requires keeping track of mappings between Logical Block Address (LBA) and Physical Block Address (PBA). As flash memory uses out-of-place update, data will become obsolete and those pages storing them will turn ``invalid'', and will be reclaimed periodically by \textit{garbage collection}. \textit{Wear leveling} is introduced to distribute programmings/erasures evenly across flash memory to prolong its lifetime. Flash cells will degrade over time, and some blocks will turn ``bad'' and can no longer reliably store data. Those blocks will be managed by \textit{bad block management}.

Note that another form of using flash memory is through a flash-specific file system, e.g., YAFFS2~\cite{YAFFS}, UBIFS~\cite{UBIFS}, F2FS~\cite{f2fs}. However, flash file systems have been used less and less in the past years. A potential reason is the flash file system requires access to the raw flash and hardware features of flash memory are visible to upper layers. On the contrary, the FTL handles the hardware features of flash memory internally, making them transparent to the upper layers, such that any systems designed for traditional block devices remain usable. In this work, we mainly focus on the FTL-based flash storage devices which dominate the flash memory storage market.

\subsection{Steganography vs. Hidden Volumes}
\label{sec:background:hidden}

To build practical PDE systems, it typically relies on either steganography or hidden volumes. A few steganographic
file systems~\cite{stegfs_IH1998,stegfs_IH1999,stegfs_ICDE2003} have been designed in the literature to hide data in regular file systems. However, all of them seem to hide sensitive data among regular file data or randomness being filled. This may result in data loss of hidden files as they may be overwritten by the regular file data. To mitigate data loss, they usually need to maintain a large amount of redundant data which will lead to inefficient use of disk space. The \textit{hidden-volume mechanism} (e.g., VeraCrypt~\cite{VeraCrypt} and TrueCrypt~\cite{truecrypt}) can mitigate the risk of data loss by intelligently placing all the hidden files towards the end of the disk. In this way, the redundant data required for data loss can be significantly reduced. A large number of existing PDE systems utilize the hidden-volume mechanism to provide deniability, including Mobiflage~\cite{mobiflage_ndss2013}, MobiHydra~\cite{mobihydra_isc2014}, MobiPluto~\cite{mobipluto_acsac2015}, HIVE~\cite{blass2014toward}, DataLair~\cite{chakraborti2017datalair}, and ECD~\cite{Zuck:2017:PHD:3102980.3102989}.

The hidden-volume mechanism works as follows: Two volumes, a public volume and a hidden volume, are created on the disk. The public volume is encrypted using a \textit{decoy key} and is placed across the entire disk. The hidden volume is encrypted using a secret \textit{true key} and is placed towards the end of the disk from a secret offset. The sensitive data being protected will be stored in the hidden volume. Note that the entire disk will be filled with random data initially. Upon being coerced by the adversary, to protect the true key, the victim can simply disclose the decoy key. Using the decoy key, the adversary can decrypt the public volume, and is not able to detect existence of the hidden volume, since he/she cannot differentiate the encrypted hidden volume from the randomness being filled initially. Corresponding to the two volumes, two modes, a \textit{public mode} and a \textit{hidden mode} are introduced to manage the public and the hidden volume, respectively. When operating in the public mode, the user can manage (e.g., read, write, process) public non-sensitive data in the public volume; when operating in the hidden mode, the user can manage (e.g., read, write, process) sensitive data in the hidden volume. 

Note that the hidden-volume mechanism can ensure deniability when the adversary can have access to the disk only once. However, if the adversary can periodically have access to the disk, the hidden-volume mechanism cannot ensure deniability, because: by comparing snapshots taken at different points of time, the adversary is able to detect changes among the ``free'' space of the public volume (caused by the hidden volume I/Os), which is not supposed to happen.

\subsection{Our Prior Research}
\label{sec:preliminary}
\minihead{Mobile PDE systems on the block-layer.} \textit{MobiPluto}~\cite{chang2015mobipluto,chang2018user} is a user-friendly PDE system, which was designed to improve usability of PDE using hidden volume mechanism. It introduces a file-system friendly PDE design which can be compatible with any block-based file systems in mobile devices. Its key idea is introducing an additional software layer between the PDE and the file system. The new layer can provide virtual volumes, on which any block-based file systems can be deployed; in addition, the layer can convert non-sequential allocation from a file system to sequential allocation in the underlying PDE to prevent over-writes on the hidden volume; finally, the layer is built using thin provisioning~\cite{thin}, a well-established tool in Linux kernel, to avoid bringing extra software components to the kernel leading to deniability compromise. 

Another PDE system on the block layer, \textit{MobiCeal}~\cite{chang2018mobiceal}, was designed to improve deniability of PDE for mobile devices. To mitigate a snapshot adversary which can obtain snapshots \textit{on the block layer} over time, \textit{MobiCeal} denies existence of hidden sensitive data using dummy writes. The key idea is: each time when writing public non-sensitive data (encrypted using decoy key), the system will perform a few artificial writes of randomness (i.e., dummy writes); the hidden sensitive data will be encrypted using true key before written to the disk.  
Deniability is ensured by using dummy writes to deny existence of the hidden writes. To further avoid deniability compromise, all the data are written to random locations of the disk~\cite{chang2018mobiceal}.

\minihead{Mobile PDE systems on the flash memory-layer.}
For the first time, \textit{DEFTL}~\cite{jia2017ccs} incorporates the hidden volume technique into the FTL (Sec.~\ref{sec:background:flash}) to ensure deniability.
In the \textit{DEFTL}, the sensitive data are stored secretly in the hidden volume. Most importantly, to prevent data written in the public volume from over-writing data written in the hidden volume, \textit{DEFTL} carefully modifies block allocation and garbage collection strategies in the FTL such that the two modes can be ``stealthily'' isolated without being known by the adversary. Specifically, the public volume will allocate flash blocks from the head of the block pool and the hidden volume will allocate flash blocks from the tail of the pool. In addition, garbage collection in the two modes will be modified as: In the public mode, garbage collection will be performed actively to fill the head of the pool; in the hidden mode, garbage collection will be performed actively to fill the tail of the pool. This can avoid that the public mode has used all the blocks in the head and starts to use the blocks in the tail, over-writing the hidden sensitive data. 

\section{Adversarial Model}
\label{sec:background:adversary}
We consider a computationally bounded snapshot adversary. The adversary can observe the victim mobile device multiple times at different points of time~\cite{blass2014toward,defy_ndss2015} (i.e., a multi-snapshot adversary). Each snapshot being taken of includes information from both storage medium and memory. The snapshot on the storage medium can be the physical image of raw NAND flash, obtainable by forensic data recovery tools~\cite{breeuwsma2007forensic}. On the contrary, most of the existing works in the literature only consider a single-snapshot adversary which can have access to the victim storage device once~\cite{mobiflage_ndss2013,mobiflage_TDSC2014,yu2014mobihydra,chang2015mobipluto,jia2017ccs,chang2018user}.

The \textit{multi-snapshot adversary} captures a lot of scenarios from real world. For example, each time when the victim enters/exits a guarded facility or crosses border, the inspector may obtain a full snapshot of his/her smart phone.
An independent journalist was reported to have all of his computers, mobile phones, and camera flash drives searched and copied when he was crossing a border, and he was inspected for seven times during five years~\cite{multiexample}. 
Additionally, an attacker can launch the ``Evil Maid'' attack that periodically sneaks into victim's hotel room and obtains multiple copies of the hard disk or memory of the targeted device. Moreover, endpoint devices such as PC and mobile phones typical use cloud-based backup systems (e.g., iCloud) that periodically snapshot the disk for data protection but employers also use the same backup data for work monitoring on their employees. The adversary is assumed to be not able to capture the device owner when she is working in the hidden mode. Otherwise, the hidden sensitive data are trivially disclosed. In addition, the adversary is assumed to be not able to have access to the random bits flipped during running of the randomization algorithm~\cite{chen2016sanitizing}.

\section{A Preliminary Design of A Secure Mobile PDE System against Multi-snapshot Adversaries}

Compared to the single-snapshot adversary, the multi-snapshot adversary is much more difficult to be combated. \textit{MobiCeal}~\cite{chang2018mobiceal} combats an adversary which can obtain multiple snapshots from the block layer by: 1) introducing a dummy write on the block layer to deny unaccountable changes caused by hidden sensitive data; and 2) writing all the data (public, dummy and hidden data) to random locations of the block layer to avoid deniability compromise~\cite{chang2018mobiceal}. \textit{MobiCeal}, however, cannot combat an adversary which can obtain multiple snapshots from the raw flash memory (Sec.~\ref{sec:background:adversary}), because: To accommodate special nature of flash memory, FTL usually follows a log-structured writing manner~\cite{Zuck:2017:PHD:3102980.3102989}, i.e., regardless how data are written on the block layer, they are written sequentially to the flash memory; thus, random writes on the block layer will become useless and deniability may be compromised by the snapshot adversary using the raw flash memory. 

To combat the multi-snapshot attacks using raw flash memory, we need to carefully modify the FTL to support PDE as follows: 1) Two modes, a public and a hidden mode, are incorporated into the FTL. When the system enters the public mode, the user can write public non-sensitive data and, when the system enters the hidden mode, the user can write hidden sensitive data. For each mode, the system should maintain a table which keeps mappings between LBAs and PBAs. Therefore, there should be two independent mapping tables in the FTL, \texttt{MAP\_PUBLIC} for the public mode and \texttt{MAP\_HIDDEN} for the hidden mode. 2) Upon performing public data writes, the FTL will perform additional dummy writes of random data. Deniability can be achieved since the encrypted sensitive data cannot be differentiated from randomness created by dummy writes and, even though the multi-snapshot adversary can notice change of randomness across flash memory, it can be denied as created by new dummy writes. One security issue here is that, rarely, there are no public data writes between the two snapshots captured by the adversary. In this case, if any hidden writes are performed between the two snapshots, the adversary will be clear that uncountable changes are caused by the hidden writes, compromising deniability. To avoid this compromise, the FTL should actively perform a few dummy writes if there are no public data writes for a long time. 3) The FTL places all the data (resulted from public, dummy and hidden writes) randomly to flash memory. The random placement technique is feasible for flash memory, because random seeks on flash memory are as efficient as sequential seeks. In addition, random placements inherently distribute data evenly among flash, naturally achieving good wear leveling~\cite{chen2016sanitizing}. Finally, public and hidden data can be easily located by the user using \texttt{MAP\_PUBLIC} and \texttt{MAP\_HIDDEN}. A few additional issues need further investigation.

\textit{First}, how can we hide the mapping table for the hidden mode (i.e., \texttt{MAP\_HIDDEN})? \texttt{MAP\_HIDDEN} can be stored encrypted (using the true key) and stored in a few random locations. The existence of flash pages storing the encrypted \texttt{MAP\_HIDDEN} can be denied as storing dummy data. These random locations can be derived from the true key only known to the hidden mode. Once the user enters the hidden mode, the FTL will use the true key to compute and locate flash pages storing \texttt{MAP\_HIDDEN}, decrypt \texttt{MAP\_HIDDEN}, and identify all flash pages storing hidden data.

\textit{Second}, how can we prevent data of the public mode from overwriting data of the hidden mode? Note that for deniability purpose, the public mode should not be aware of existence of the hidden mode; otherwise, the adversary who can have access to the public mode will trivially know the existence of hidden sensitive data. To address this issue, a global bitmap should be introduced to the FTL to keep track of usage of flash pages. Specifically, once a flash page is used (by either public, dummy or hidden data), the page will be marked as used in the global bitmap. This would not be vulnerable to the multi-snapshot adversary, since the flash pages used by the hidden data can be denied as used by the dummy data. Another issue is how to store and maintain the global bitmap. Storing the bitmap in the flash memory may be problematic, since flash memory is update unfriendly (Sec.~\ref{sec:background:flash}), but the bitmap needs to be updated frequently in order to keep track of page usage. Storing the bitmap in memory can solve the problem. However, information storing in memory will suffer from loss upon power failures. A preliminary solution could be: when the mobile device is powered off normally, the bitmap in memory will be committed to flash memory; when the mobile device encounters a sudden power failure, the FTL will perform a full disk scan to reconstruct the bitmap. The user should be involved and provide both the decoy key and the true key in order to localize the public and hidden data, reconstructing the portion of bitmap for them. There is no need to reconstruct the portion of bitmap for the dummy data and the corresponding flash pages can be reused. Since the power failure is a rare event, the user involvement will not create usability issues.

\textit{Third}, how can we prevent garbage collection (Sec.~\ref{sec:background:flash}) from reclaiming space occupied by hidden sensitive data and \texttt{MAP\_HIDDEN}? Without garbage collection, dummy data will fill the entire disk soon. Therefore, space occupied by dummy data needs to be reclaimed periodically. Note that during each garbage collection, the FTL should only reclaim a random portion of flash pages occupied by dummy data to avoid deniability compromise. The challenge is, the public mode cannot differentiate randomness created by the hidden writes from that created by the dummy writes. Having observed that the hidden mode knows where are the actual dummy data, a promising strategy could be using a ``lazy'' garbage collection in the public mode while using an active garbage collection in the hidden mode. Specifically, the public mode will not reclaim space occupied by dummy data if flash memory load factor (i.e., percentage of the entire space being used) does not exceed a large threshold and, when the user enters the hidden mode, the FTL will actively reclaim flash blocks occupied by dummy data and invalid data. To avoid significantly impact performance of the hidden mode, the garbage collection could be performed during idle time of the hidden mode.

\section{Conclusion}
In this work, we propose a preliminary mobile PDE system design which aims to defend against multi-snapshot adversaries. A few fundamental limitations of such a design are: 1) Dummy writes could cause significant overhead to both I/O and storage, and how to bound the amount of dummy writes without sacrificing security is a challenging problem which deserves further investigation; 2) The adversary should not have access to the random bits flipped during the running of the algorithm and, cannot capture the device owner while working in the hidden mode. These are all impractical assumptions which need to be further relaxed. Our on-going work will address the aforementioned limitations.

\section*{Acknowledgment}
We would like to thank Fengwei Zhang for the initial discussion of the design during 2018. This work was supported by National Science Foundation under award number 1928349-CNS and 1938130-CNS. Any opinions, findings, and conclusions in this document are those of the author(s) and do not necessarily reflect the views of the National Science Foundation.

\thispagestyle{empty}
\bibliographystyle{plain}
\bibliography{reference}

\begin{thebibliography}{10}

\bibitem{f2fs}
F2fs.
\newblock \url{https://www.kernel.org/doc/Documentation/filesystems/f2fs.txt}.

\bibitem{UBIFS}
Ubifs.
\newblock \url{http://www.linux-mtd.infradead.org/doc/ubifs.html}.

\bibitem{Rubberhose}
{ Julian Assange, Ralf P. Weinmann and Suelette Dreyfus}.
\newblock {Rubberhose Filesystem}.
\newblock {\em \url{https://github.com/sporkexec/rubberhose}}, 2000.

\bibitem{stegfs_IH1998}
Ross Anderson, Roger Needham, and Adi Shamir.
\newblock The steganographic file system.
\newblock In {\em International Workshop on Information Hiding}, pages 73--82.
  Springer, 1998.

\bibitem{blass2014toward}
Erik-Oliver Blass, Travis Mayberry, Guevara Noubir, and Kaan Onarlioglu.
\newblock Toward robust hidden volumes using write-only oblivious ram.
\newblock In {\em Proceedings of the 2014 ACM SIGSAC Conference on Computer and
  Communications Security}, pages 203--214. ACM, 2014.

\bibitem{breeuwsma2007forensic}
Marcel Breeuwsma, Martien De~Jongh, Coert Klaver, Ronald Van Der~Knijff, and
  Mark Roeloffs.
\newblock Forensic data recovery from flash memory.
\newblock {\em Small Scale Digital Device Forensics Journal}, 1(1):1--17, 2007.

\bibitem{chakraborti2017datalair}
Anrin Chakraborti, Chen Chen, and Radu Sion.
\newblock Datalair: Efficient block storage with plausible deniability against
  multi-snapshot adversaries.
\newblock {\em Proceedings on Privacy Enhancing Technologies}, 3:175--193,
  2017.

\bibitem{chang2018user}
Bing Chang, Yao Cheng, Bo~Chen, Fengwei Zhang, Wen-Tao Zhu, Yingjiu Li, and
  Zhan Wang.
\newblock User-friendly deniable storage for mobile devices.
\newblock {\em Computers \& Security}, 72:163--174, 2018.

\bibitem{mobipluto_acsac2015}
Bing Chang, Zhan Wang, Bo~Chen, and Fengwei Zhang.
\newblock Mobipluto: File system friendly deniable storage for mobile devices.
\newblock In {\em Proceedings of the 31st Annual Computer Security Applications
  Conference}, pages 381--390. ACM, 2015.

\bibitem{chang2015mobipluto}
Bing Chang, Zhan Wang, Bo~Chen, and Fengwei Zhang.
\newblock Mobipluto: File system friendly deniable storage for mobile devices.
\newblock In {\em Proceedings of the 31st Annual Computer Security Applications
  Conference}, pages 381--390. ACM, 2015.

\bibitem{chang2018mobiceal}
Bing Chang, Fengwei Zhang, Bo~Chen, Yingjiu Li, Wen-Tao Zhu, Yangguang Tian,
  Zhan Wang, and Albert Ching.
\newblock Mobiceal: Towards secure and practical plausibly deniable encryption
  on mobile devices.
\newblock In {\em 2018 48th Annual IEEE/IFIP International Conference on
  Dependable Systems and Networks (DSN)}, pages 454--465. IEEE, 2018.

\bibitem{chen2016sanitizing}
Bo~Chen, Shijie Jia, Luning Xia, and Peng Liu.
\newblock Sanitizing data is not enough!: towards sanitizing structural
  artifacts in flash media.
\newblock In {\em Proceedings of the 32nd Annual Conference on Computer
  Security Applications}, pages 496--507. ACM, 2016.

\bibitem{VeraCrypt}
CodePlex.
\newblock Veracrypt ssd.
\newblock \url{https://veracrypt.codeplex.com/}, 2017.

\bibitem{multiexample}
Amy Goodman.
\newblock Dn! exclusive: Authorities search and copy u.s. journalist’s notes,
  computer and cameras after returning from haiti.
\newblock
  \url{https://www.democracynow.org/2011/2/15/exclusive_authorities_search_and_copy_us},
  2011.

\bibitem{gupta2009dftl}
Aayush Gupta, Youngjae Kim, and Bhuvan Urgaonkar.
\newblock {\em DFTL: a flash translation layer employing demand-based selective
  caching of page-level address mappings}, volume~44.
\newblock ACM, 2009.

\bibitem{jia2017ccs}
Shijie Jia, Luning Xia, Bo~Chen, and Peng Liu.
\newblock Deftl: Implementing plausibly deniable encryption in flash
  translation layer.
\newblock In {\em Proceedings of the 24th ACM conference on Computer and
  communications security}. ACM, 2017.

\bibitem{kang2006superblock}
Jeong-Uk Kang, Heeseung Jo, Jin-Soo Kim, and Joonwon Lee.
\newblock A superblock-based flash translation layer for nand flash memory.
\newblock In {\em Proceedings of the 6th ACM \& IEEE International conference
  on Embedded software}, pages 161--170. ACM, 2006.

\bibitem{kim2002space}
Jesung Kim, Jong~Min Kim, Sam~H Noh, Sang~Lyul Min, and Yookun Cho.
\newblock A space-efficient flash translation layer for compactflash systems.
\newblock {\em IEEE Transactions on Consumer Electronics}, 48(2):366--375,
  2002.

\bibitem{lee2007log}
Sang-Won Lee, Dong-Joo Park, Tae-Sun Chung, Dong-Ho Lee, Sangwon Park, and
  Ha-Joo Song.
\newblock A log buffer-based flash translation layer using fully-associative
  sector translation.
\newblock {\em ACM Transactions on Embedded Computing Systems (TECS)}, 6(3):18,
  2007.

\bibitem{stegfs_IH1999}
Andrew~D McDonald and Markus~G Kuhn.
\newblock Stegfs: A steganographic file system for linux.
\newblock In {\em Information Hiding}, pages 463--477. Springer, 2000.

\bibitem{stegfs_ICDE2003}
HweeHwa Pang, K-L Tan, and Xuan Zhou.
\newblock Stegfs: A steganographic file system.
\newblock In {\em Data Engineering, 2003. Proceedings. 19th International
  Conference on}, pages 657--667. IEEE, 2003.

\bibitem{park2008reconfigurable}
Chanik Park, Wonmoon Cheon, Jeonguk Kang, Kangho Roh, Wonhee Cho, and Jin-Soo
  Kim.
\newblock A reconfigurable ftl (flash translation layer) architecture for nand
  flash-based applications.
\newblock {\em ACM Transactions on Embedded Computing Systems (TECS)}, 7(4):38,
  2008.

\bibitem{defy_ndss2015}
Timothy~M Peters, Mark~A Gondree, and Zachary~NJ Peterson.
\newblock D{E}{F}{Y}: A deniable, encrypted file system for log-structured
  storage.
\newblock In {\em 22th Annual Network and Distributed System Security
  Symposium, NDSS}, 2015.

\bibitem{mobiflage_ndss2013}
Adam Skillen and Mohammad Mannan.
\newblock On implementing deniable storage encryption for mobile devices.
\newblock In {\em 20th Annual Network and Distributed System Security
  Symposium, {NDSS} 2013, San Diego, California, USA, February 24-27}, 2013.

\bibitem{mobiflage_TDSC2014}
Adam Skillen and Mohammad Mannan.
\newblock Mobiflage: Deniable storage encryptionfor mobile devices.
\newblock {\em IEEE Transactions on Dependable and Secure Computing},
  11(3):224--237, 2014.

\bibitem{androidfde}
Source.
\newblock Android full disk encryption.
\newblock \url{https://source.android.com/security/encryption/}, 2016.

\bibitem{thin}
Joe Thornber.
\newblock Thin provisioning tools.
\newblock \url{https://github.com/jthornber/thin-provisioning-tools}.

\bibitem{truecrypt}
{TrueCrypt}.
\newblock Free open source on-the-fly disk encryption software.version 7.1a.
\newblock {\em Project website: \url{http://www.truecrypt.org/}}, 2012.

\bibitem{YAFFS}
Yaffs.
\newblock Yaffs.
\newblock \url{http://www.yaffs.net/}, 2002.

\bibitem{mobihydra_isc2014}
Xingjie Yu, Bo~Chen, Zhan Wang, Bing Chang, Wen~Tao Zhu, and Jiwu Jing.
\newblock Mobihydra: Pragmatic and multi-level plausibly deniable encryption
  storage for mobile devices.
\newblock In {\em Information Security - 17th International Conference, {ISC}
  2014, Hong Kong, China, October 12-14, 2014. Proceedings}, pages 555--567,
  2014.

\bibitem{yu2014mobihydra}
Xingjie Yu, Bo~Chen, Zhan Wang, Bing Chang, Wen~Tao Zhu, and Jiwu Jing.
\newblock Mobihydra: Pragmatic and multi-level plausibly deniable encryption
  storage for mobile devices.
\newblock In {\em International Conference on Information Security}, pages
  555--567. Springer, 2014.

\bibitem{Zuck:2017:PHD:3102980.3102989}
Aviad Zuck, Udi Shriki, Donald~E. Porter, and Dan Tsafrir.
\newblock Preserving hidden data with an ever-changing disk.
\newblock In {\em Proceedings of the 16th Workshop on Hot Topics in Operating
  Systems}, HotOS '17, pages 50--55, New York, NY, USA, 2017. ACM.

\end{thebibliography}

\end{document}